\documentclass[twocolumn,preprintnumbers,amsmath,amssymb,superscriptaddress]{revtex4}
\usepackage{subfigure}
\usepackage{graphicx}
\usepackage{dcolumn}
\usepackage{bm}
\usepackage{color}
\usepackage{url}
\usepackage{hyperref}
\usepackage[utf8]{inputenc}
\usepackage{float}

\def \equi#1{\mathrel{\mathop{\kern 0pt\sim}\limits_{#1}}} 

\begin{document}

\title{Optimally Frugal Foraging}

\author{O. B\'enichou} \affiliation{Laboratoire de Physique Th\'eorique de la
  Mati\`ere Condens\'ee (UMR CNRS 7600), Universit\'e Pierre et Marie Curie,
  4 Place Jussieu, 75252 Paris Cedex France}
  
\author{U. Bhat} \affiliation{Santa Fe Institute, 1399 Hyde Park Road, Santa Fe,
  New Mexico 87501, USA} \affiliation{Department of Physics, Boston University,
  Boston, MA 02215, USA} \affiliation{School of Natural Sciences, University of
  California at Merced, Merced, CA 95343, USA}

\author{P. L. Krapivsky}
\affiliation{Department of Physics, Boston University, Boston, MA 02215, USA}

\author{S. Redner} \affiliation{Santa Fe Institute, 1399 Hyde Park Road, Santa
  Fe, New Mexico 87501, USA}

\begin{abstract}

   We introduce the \emph{frugal foraging} model in which a forager performs a
  discrete-time random walk on a lattice, where each site initially contains
  $\mathcal{S}$ food units.  The forager metabolizes one unit of food at each
  step and starves to death when it last ate $\mathcal{S}$ steps in the past.
  Whenever the forager decides to eat, it consumes all food at its current
  site and this site remains empty (no food replenishment).  The crucial
  property of the forager is that it is \emph{frugal} and eats only when
  encountering food within at most $k$ steps of starvation.  We compute the
  average lifetime analytically as a function of frugality threshold and show
  that there exists an optimal strategy, namely, a frugality threshold $k^*$
  that maximizes the forager lifetime.
 
\end{abstract}

\maketitle

Foraging is a fundamental ecological process that has sparked much research
(see, e.g.,
\cite{charnov1976optimal,Stephens-DW:1986,Bell:1991,OBrien:1990}).  Theories
of foraging have attempted to determine strategies for a forager to maximize
food consumption.  Such strategies balance the interplay between
exploitation, where a forager consumes food in a current search domain, and
exploration, where a forager moves to potentially richer domains.  This
dichotomy underlies many phenomena for which statistical physics ideas have
been fruitful, including management of
firms~\cite{march1991exploration,gueudre2014explore}, the multiarm bandit
problem~\cite{robbins1952some,gittins1979bandit}, the secretary
problem~\cite{ferguson1989solved}, Feynman's restaurant problem~\cite{FRP2},
and human memory~\cite{hills2012optimal,abbott2015random}.

Typically these optimization problems do not account for depletion of
resources.  The \emph{starving random
  walk}~\cite{PhysRevLett.113.238101,benichou2016role,Chupeau:2017}
explicitly accounts for this basic coupling between forager motion and
depletion.  In this model, a forager random walks on a lattice in which each
site initially contains $\mathcal{S}$ food units.  When a forager lands on a
food-containing site, all the food there is consumed and the forager is fully
sated.  The forager metabolizes one unit of food at each step so that it
starves to death when it last ate $\mathcal{S}$ time steps in the past.  When
the forager lands on an empty site, it comes one time unit closer to
starvation.  Because food is not replenished, the forager is doomed to
starve, with a non-trivial dependence of lifetime on its metabolic capacity
$\mathcal{S}$ and the spatial dimension
$d$~\cite{PhysRevLett.113.238101,benichou2016role}.

\begin{figure}[ht]
\begin{center}
\includegraphics[width=0.4\textwidth]{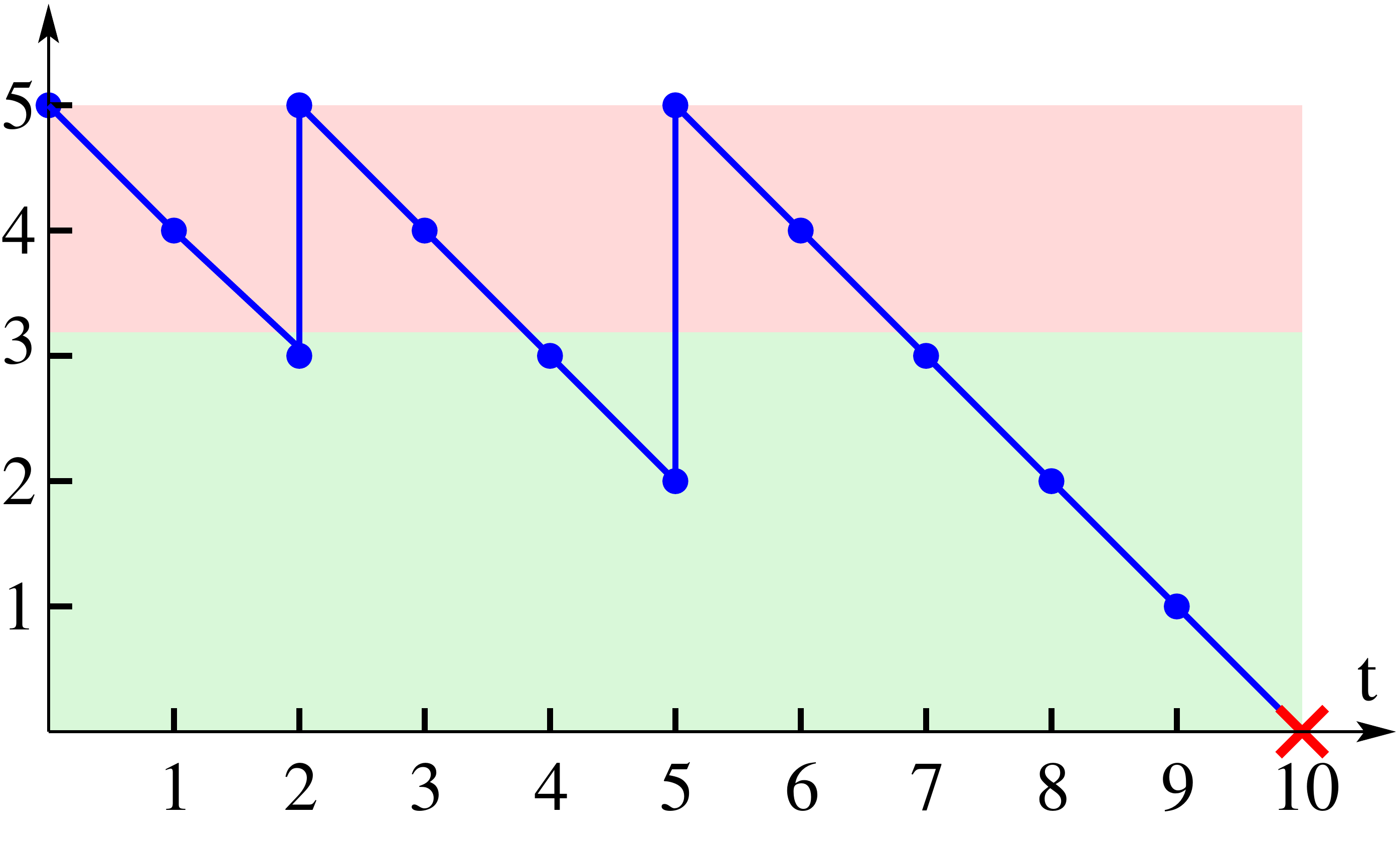}
\caption{The nutritional state of a frugal forager as a function of time for
  the case $\mathcal{S}=5$ and frugality threshold $k=3$.  In the red zone,
  the forager does not eat, even when it encounters food, while in the green
  zone the forager eats whenever it encounters food.  This forager starves at
  $t=10$.}
\label{fig:evol}
  \end{center}
\end{figure}

In the starving random walk, the forager mindlessly eats whenever food is
encountered.  Is it possible that the forager can live longer with a
different consumption strategy?  By incorporating the attribute of
\emph{frugality}, in which the forager eats only when it is nutritionally
depleted below a specified level (Fig.~\ref{fig:evol}), we show that the
average lifetime of a forager is dramatically increased.  This frugality
mimics ecological foraging, where foragers reduce their activity when
satiated and resume foraging only when sufficiently depleted; parallel
behavior occurs in predatory
animals~\cite{mills1982satiation,dill1983adaptive,sih1984optimal,jung1985effects,croy1991influence,candler1995flight,ben2001satiation,sass2002effects,mccue2012comparative,mysterud2017evolutionary}.
While delaying consumption might seem a risky survival strategy, we will show
that: (i) frugality typically increases the forager lifetime and (ii) the
lifetime is maximized at an optimal frugality.

\emph{Model.}  The frugal forager starts at the origin in a food paradise,
where each lattice site initially contains $\mathcal{S}$ food units.  The
forager immediately eats the food at the origin so it begins fully satiated.
The frugal forager performs a lattice random walk and its dynamics is
identical to that of the starving random
walk~\cite{PhysRevLett.113.238101,benichou2016role} (which here we term the
normal forager), \emph{except} that the frugal forager can eat only when it
encounters a food-containing site within $k$ hops---the frugality
threshold---of starving, with $0\leq k\leq\mathcal{S}-1$
(Fig.~\ref{fig:evol}).  The case $k=\mathcal{S}-1$ corresponding to the
normal forager that eats anytime it encounters food.  Upon eating, the
forager returns to a fully satiated state (so that up to $k$ food units are
``wasted'' by this consumption rule). This foraging process is inherently
non-Markovian because it depends on all of a forager's previous encounters
with food, i.e., the times between visits to distinct sites of a random
walk~\cite{polya:1921,doi:10.1063/1.1704269,Weiss:1994,Hughes:1995}.

\begin{figure}[ht]
\begin{center}
\subfigure[]{  \includegraphics[width=0.425\textwidth]{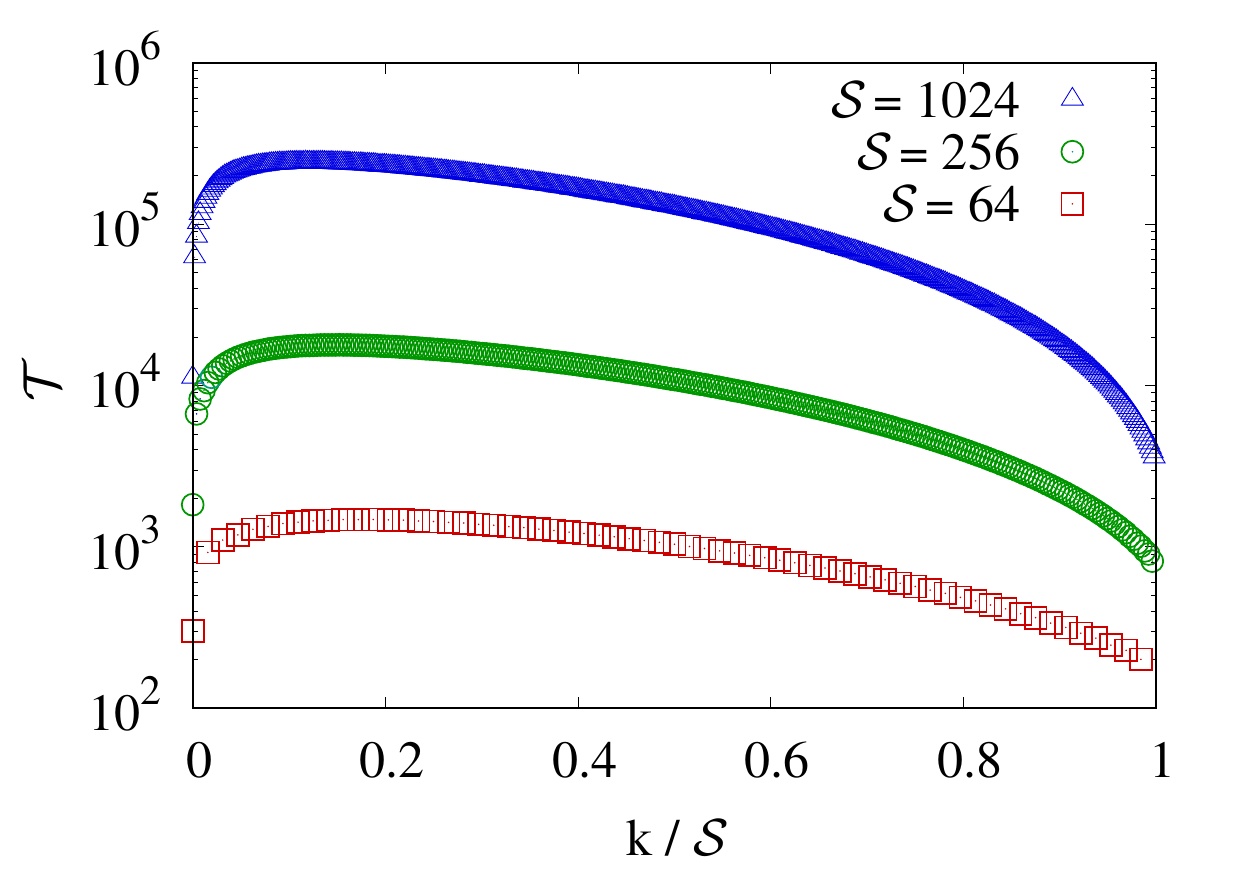}}
\subfigure[]{  \includegraphics[width=0.425\textwidth]{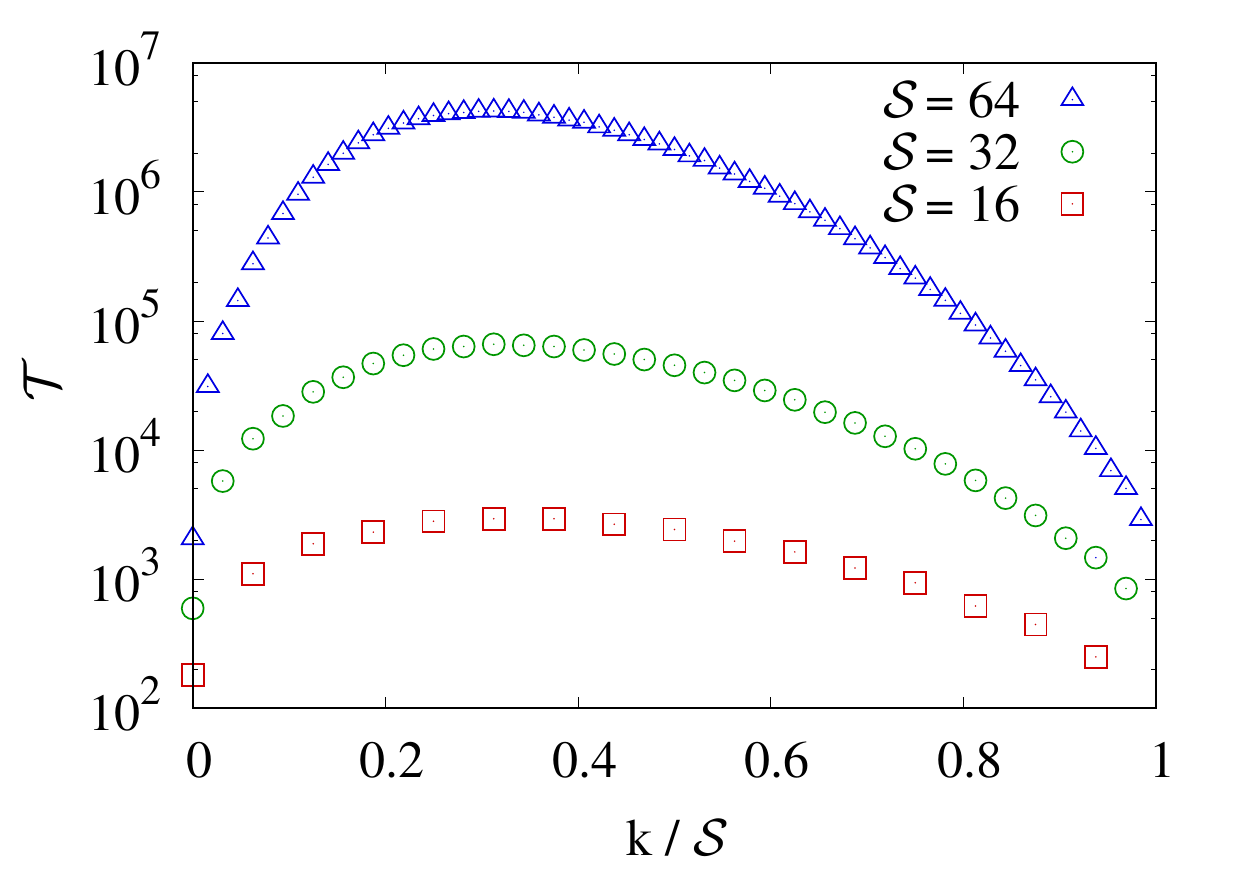}}
\caption{Simulation results for dependence of the forager lifetime on
  frugality threshold $k$ for various metabolic capacities $\mathcal{S}$ in
  (a) $d=1$ and (b) $d=2$.  The case $k=\mathcal{S}-1$ corresponds to a
  normal forager (starving random walk).}
\label{fig:t}
  \end{center}
\end{figure}

\emph{Simulations.}  We find that the average forager lifetime is maximized
at a optimal thresholds $k^*$ in both $d=1$ and $d=2$ (Figure~\ref{fig:t}).
For $d=3$, a qualitatively similar optimization arises.  Thus optimal
frugality is generally advantageous.  It is also worth noting that as the
metabolic capacity $\mathcal{S}$ increases, which correlates with larger body
size, a forager maximizes its lifetime by becoming progressively more frugal;
that is, $k^*$ is a decreasing function of $\mathcal{S}$.

To understand the optimization of the forager lifetime, we now investigate
the tractable cases $k\ll\mathcal{S}$, where we will show that the dependence
of the lifetime $\mathcal{T}$ on $\mathcal{S}$ steepens for increasing $k$
and also grows more quickly than for the normal forager, where
$k=\mathcal{S}-1$.  These two facts mandate that there must exist an optimum
frugality $k^*$, where the $\mathcal{S}$ dependence of the lifetime is the
fastest.

\emph{Maximally frugal forager.}  To analytically show that always consuming
when resources are encountered is a suboptimal survival strategy, we treat
the extreme case of a maximally frugal forager that can only eat at the
instant of starvation, i.e., $k=0$.  To continue to survive, the forager must
therefore be at previously unvisited sites at times $\mathcal{S}$,
$2\mathcal{S}$, $3\mathcal{S}$, etc.  (For simplicity, we consider even
$\mathcal{S}$ and hypercubic lattices.)~ If the forager lands on a previously
visited site (where food was consumed) at time $m\mathcal{S}$, with $m$ an
integer, starvation immediately occurs and the forager lifetime is
$m\mathcal{S}$.  It is worth noticing that we can also think of the maximally
frugal forager as a self-avoiding flight~\cite{HN85,LND87,H11}, in which each
step of the flight is determined by the displacement of a nearest-neighbor
random walk of $\mathcal{S}$ steps and the flight dies whenever it lands on a
previously visited site.  The result \eqref{general} for the forager survival
probability that we will derive also describes to the survival probability of
a self-avoiding flight whose return probability at a single step is given by
\eqref{Rj}.

We first show that the survival probability decays exponentially in time in
any dimension.  Define $S_m$ as the probability that a maximally frugal
forager survives until time $m\mathcal{S}$, and $R_m$ as the probability for
a pure random walk to return to its starting point at time $m\mathcal{S}$.
The forager survives its first potential starvation event at time
$\mathcal{S}$ with probability $S_1=1-R_1$.  We obtain an upper bound for the
survival probability at time $2\mathcal{S}$ by demanding that the forager
hops to a different site from where it was at time $\mathcal{S}$.  The
probability for this event is again $S_1$.  Because we have not included the
possibility that the forager has returned to the origin at time
$2\mathcal{S}$, the true survival probability will be smaller still.
Therefore $S_2\leq (S_1)^2$.  Continuing this reasoning gives
$S_m\le(S_1)^m$, a result that is valid for any $\mathcal{S}$.

We obtain a lower bound by noting that the forager is sure to survive if its
position always has a positive increment in a single coordinate direction
between times $m\mathcal{S}$ and $(m+1)\mathcal{S}$.  Let $Q_1(d)$ be the
probability that a random walk has a single coordinate equal to zero at time
$\mathcal{S}$ in $d$ dimensions.  Then the probability that a single
coordinate has increased is $\frac{1}{2}\big(1-Q_1(d)\big)$.  We therefore
have the bounds
\begin{equation}
\label{bounds}
2d\,\,(\tfrac{1}{2})^m\big[1-Q_1(d)\big]^m\le S_m\le (1-R_{1})^m\,,
\end{equation}
so that $S_m$ asymptotically decays exponentially in $m$ and,
correspondingly, exponentially in time.  However, we will show the mean
lifetime of the forager in low spatial dimension is controlled by an
intermediate regime for large $\mathcal{S}$, where the survival probability
decays as a super exponential in $m$ for $d\leq 2$.

We start by deriving an exact recurrence that is satisfied by the survival
probability of a maximally frugal forager and then give explicit results for
$\mathcal{S}\to\infty$.  Formally, the survival probability is given by
\begin{subequations}
\begin{equation}
  S_m={\rm Pr}\{\Delta_m=1,\Delta_{m-1}=1,\cdots,\Delta_1=1\}\,,
\end{equation}
where $\Delta_j$ is the random variable that equals 1 if a new site is
visited at time $j$ and equals 0 otherwise.  The survival probability
satisfies the recursion
\begin{align}
  \label{recurrence1}
  S_m&={\rm Pr}\{\Delta_m=1|\Delta_{m-1}=1,\cdots,\Delta_1=1\}\times\nonumber\\
&~~~~~~~~~~ {\rm Pr}\{\Delta_{m-1}=1,\cdots,\Delta_1=1\}\,,\nonumber\\
&={\rm Pr}\{\Delta_m=1|\Delta_{m-1}=1,\cdots,\Delta_1=1\}\,S_{m-1}\,.
\end{align}
\end{subequations}

To obtain explicit results from this exact recurrence, we make the
approximation
\begin{equation}
\label{approx}
{\rm Pr}\{\Delta_m\!=\!1|\Delta_{m-1}\!=\!1,\cdots,\Delta_1\!=\!1\}\simeq{\rm Pr}\{\Delta_m\!=\!1\}\,;
\end{equation}
i.e., correlations with past events are ignored.  To determine
${\rm Pr}\{\Delta_m=1\}$, it is useful to introduce its generating function,
which is known to be~\cite{Hughes:1995}
\begin{equation}
  \label{Dz}
  \Delta(z)\equiv \sum_{n\geq 0} {\rm Pr}\{\Delta_n=1\} z^n=-1+\frac{1}{(1-z)R(z)}\,.
\end{equation}
Here $R(z)\equiv \sum_{n\geq 0} R_nz^n$ is the generating function for the
return probability $R_n$ at the $(n\mathcal{S})^{\rm th}$ step, which has the
asymptotic behavior~\cite{Hughes:1995}:
\begin{equation}
  \label{Rj}
  R_j\simeq 2\left[{d}/(2\pi j \mathcal{S})\right]^{d/2}\,.
\end{equation}

For $\mathcal{S}\gg 1$, which we assume throughout, $1/R(z)$ has the
expansion
\begin{align*}
&1-R_1z+\big(R_1^2-R_2\big)z^2-\big(R_1^3-2R_1R_2+R_3)z^3+\cdots\,,\nonumber\\
&\simeq1-\sum_{j\geq 1}R_jz^j\,.
\end{align*}
In the second line, powers of $R_j$ greater than 1 are negligible compared to
linear terms for $\mathcal{S}\to\infty$ because $R_j$ asymptotically scales
as $\mathcal{S}^{-d/2}$

Substituting the above expansion for $1/R(z)$ in Eq.~\eqref{Dz}, we obtain the
series for $\Delta(z)$, from which we read off
${\rm Pr}\{\Delta_m=1\}=1-\sum_{1\leq j\leq m}R_j$.  Using this expression in
Eqs.~\eqref{recurrence1} and \eqref{approx} gives
\begin{equation}
\label{general}
S_m\simeq \exp\bigg(-\sum_{\ell=1}^m\sum_{j=1}^\ell R_j\bigg)\,.
\end{equation}

\begin{figure}[h]
\begin{center}
\includegraphics[width=0.425\textwidth]{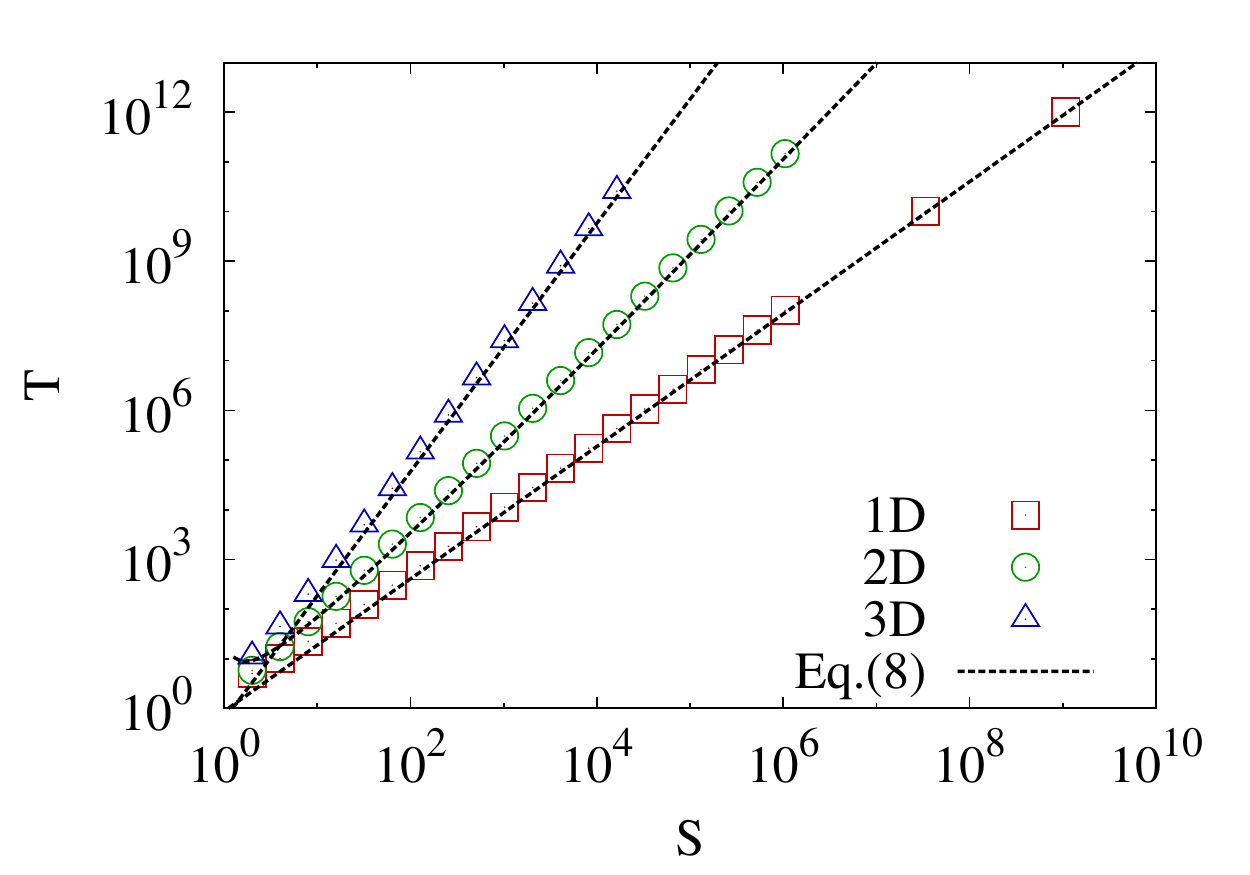}
\caption{Simulation results for the lifetime $\mathcal{T}$ of the maximally
  frugal forager versus $\mathcal{S}$.}
\label{fig:eq8}
  \end{center}
\end{figure}

Substituting $R_j$ from \eqref{Rj} in \eqref{general}, we obtain
\begin{equation}
  \label{lnSm}
-\ln S_m\simeq
\begin{cases}
{\displaystyle  \sqrt{\frac{32 m^3}{9\pi}}\,\, \mathcal{S}^{-1/2}} & \qquad d=1\,,\\[3mm]
{\displaystyle  \frac{2 m \ln m}{\pi}\,\, \mathcal{S}^{-1}} & \qquad d=2\,,\\[3.5mm]
{\displaystyle  m A_d \,\,\mathcal{S}^{-d/2}} & \qquad d>2\,,
\end{cases}
\end{equation}
where $A_d\equiv 2\zeta(d/2)(d/2\pi)^{d/2}$.  Note that the lower bound
\eqref{bounds} for the survival probability imposes the constraint that
\eqref{lnSm} cannot hold when $m\ge\mathcal{S} $ in $d=1$, and when
$m\ge e^{\mathcal{S}}$ in $d=2$.  Imposing this constraint, the average
number of ``generations'' that forager survives in $d=1$ can be found from
\begin{align*}
  \langle m \rangle &\!=\!\sum_{m\geq 1} S_m
   \simeq \int_0^{\beta \mathcal{S}}
    \!\!\!\!e^{-\alpha\left({m^3}/{\mathcal{S}}\right)^{1/2}}{\rm d}m
 +\int_{\beta \mathcal{S}}^\infty \!\!\!e^{-\gamma m}\,{\rm d}m\,,\nonumber\\
\end{align*}
with $\alpha=\sqrt{32/9\pi}$, while $\beta$ and $\gamma$ are constants of
order 1 that are irrelevant for large $\mathcal{S}$.  The integral over the
finite range dominates.  Computing this integral and performing similar
calculations in higher dimensions leads to
\begin{equation}
  \label{T}
\mathcal{T} \simeq
\begin{cases}
{\displaystyle \Gamma(2/3)\left(\frac{\pi}{12}\right)^{1/3}\mathcal{S}^{4/3}}& \qquad d=1\,,\\[3mm]
{\displaystyle\frac{\pi}{2\ln \mathcal{S}}\,\,\mathcal{S}^2} & \qquad d=2\,,\\[2mm]
{\displaystyle \frac{1}{A_d}\,\, \mathcal{S}^{1+d/2}} & \qquad d>2\,.
\end{cases}
\end{equation}
Our numerical results for $d=1$, 2, and 3 agree with the predictions of
Eq.~\eqref{T} (Fig.~\ref{fig:eq8}).

From \eqref{T}, the maximally frugal forager in $d=1$ lives longer than the
normal forager, whose lifetime is
$\mathcal{T}\sim
\mathcal{S}$~\cite{PhysRevLett.113.238101,benichou2016role}.  Moreover,
the maximally frugal forager consumes $\mathcal{S}\langle m\rangle \sim S^{4/3}$ units
of food over its lifetime, while the lifetime consumption of the normal
forager is $\mathcal{S}^{3/2}$.  Despite living a factor of
$\mathcal{S}^{1/3}$ longer, the frugal forager asymptotically consumes a
factor $\mathcal{S}^{-1/6}$ less resources.  Thus extreme frugality in one
dimension leads to a longer lifetime and less resource
consumption.

\emph{General frugality.}  We now treat the case of frugality parameter
$k>0$, where the forager can eat up to $k$ time steps before starvation, and
show that the average lifetime has an optimum with respect to $k$.  For the
case $k=1$, the forager starves when it hopes onto a doublet of two
consecutive empty (previously visited) sites and remain within the doublet at
the next step.  In the spirit of our argument that led to
Eq.~\eqref{general}, the probability to survive $m$ generations satisfies
\begin{equation}
  \label{Sm2}
S_m(k=1)\simeq \exp\left[-\sum_{\ell=1}^m\Big(\sum_{j=1}^\ell R_j\Big)^2\right]\,.
\end{equation}
Using the expression for $R_j$ given in \eqref{Rj}, we find
\begin{equation}
  \label{lnSm1}
-\ln S_m(k\!=\!1)\simeq
\begin{cases}
{\displaystyle \frac{4m^2}{\pi}\,\mathcal{S}^{-1}} &\quad d=1\,,\\[1.5mm]
{\displaystyle \frac{4 m \ln^2 m}{\pi^2}\,\, \mathcal{S}^{-2} } &\quad d=2\,,\\[3mm]
{\displaystyle mA_d^2\,\, \mathcal{S}^{-d}} &\quad d>2\,.
\end{cases}
\end{equation}
By following the steps parallel to those that gave Eq.~\eqref{T}, the average
lifetime $\mathcal{T}$ for $k=1$ is
\begin{equation}
   \label{T1}
 \mathcal{T}(k=1) \simeq
\begin{cases}
 {\displaystyle \frac{\pi}{\sqrt{8}}\,\,\mathcal{S}^{3/2}} & \qquad d=1\,, \\[2mm]
 {\displaystyle \left(\frac{\pi}{4 \ln  \mathcal{S}}\right)^2\mathcal{S}^3} & \qquad d=2\,,\\[3mm]
 {\displaystyle \frac{1}{A_d^2}\, \mathcal{S}^{1+d}} & \qquad d>2\,.
\end{cases}
\end{equation}
The lifetime for $k=1$ exceeds that of the maximally frugal forager.  Because
the lifetime of the normal forager ($k=\mathcal{S}-1$) is shorter than that
of the maximally frugal forager ($k=0$), there must be an intermediate
frugality value $k^*$ that maximizes the lifetime.

Finally, we extend our approach to general frugality threshold $k$, with
$k\ll\mathcal{S}$.  For a forager to starve in one dimension, it first has to
be metabolically depleted to its frugality threshold, then hop to the
interior of a gap of consecutive previously visited sites, and finally make
$k$ subsequent hops within this gap.  The average length of a gap that will
trap the forager is simply $\langle N(k)\rangle$, the mean number of distinct
sites visited by a random walk of $k$ steps~\cite{Weiss:1994,Hughes:1995}.
Following the same reasoning as that applied for the case $k=1$, we obtain
\begin{equation}
  \label{Smk}
  S_m(k)\simeq \exp\left[-\sum_{\ell=1}^m\Big(\sum_{j=1}^\ell R_j\Big)^{\langle N(k)\rangle}\right]\,.
\end{equation}
Here we make the uncontrolled approximation that the survival probability
averaged over all random-walk trajectories can be expressed averaging the
number of distinct sites visited in the exponent of the above expression.

Substituting the return probability \eqref{Rj} for nearest-neighbor random
walks into \eqref{Smk}, the leading behavior of the survival probability is
\begin{equation}
  \label{lnSmk}
-\ln S_m(k)\sim
\begin{cases}
{m^{1+\langle N(k)\rangle/2}}/{\mathcal{S}^{\langle N(k)\rangle/2}} &\quad d=1\,,\\[2mm]
{m (\ln m)^{\langle N(k)\rangle}}/{\mathcal{S}^{\langle N(k)\rangle}} &\quad d=2\,,\\[2mm]
{m}/{\mathcal{S}^{d\langle N(k)\rangle/2}} &\quad d>2\,.
\end{cases}
\end{equation}
From these results, we obtain the lifetime
\begin{equation}
  \label{Tgen}
\mathcal{T}(k) \sim
\begin{cases}
  \mathcal{S}^{{(2\langle N(k)\rangle+2)}/{(\langle N(k)\rangle+2)}} &\qquad d=1\,,\\[2mm]
{\mathcal{S}^{\langle N(k)\rangle+1}}/{(\ln\mathcal{S})^{\langle N(k)\rangle}  }& \qquad d=2\,,\\[2mm]
\mathcal{S}^{1+{d\langle N(k)\rangle}/{2}} & \qquad d>2\,,
\end{cases}
\end{equation}
with the asymptotic behavior of $\langle N(k)\rangle$ for $k\gg 1$ given
by~\cite{Weiss:1994,Hughes:1995}
$\sqrt{8k/\pi}$ ($d=1$), $\pi k/\ln k$ ($d=2$), and $k/R(1)$ ($d=3$).

Because of the uncontrolled nature of the approximation in \eqref{Smk}, one
should not anticipate that our prediction for the dependence of $\mathcal{T}$
on $\mathcal{S}$ for different frugality thresholds $k$ will match simulation
results quantitatively.  However, these two results are gratifyingly close in
spite of the crudeness of our approach (Table~\ref{table:exp}).

\begin{table}[H]
  \caption{Comparison between simulation results for the exponent $\tau$ in
    $\mathcal{T}\sim\mathcal{S}^\tau$ in one dimension (top row) and our
    analytical predictions (bottom row): Eq.~\eqref{T} for $k=0$,
    Eq.~\eqref{T1} for $k=1$, and then the first line of \eqref{Tgen}.}
  \centering 
  \vspace{3mm}
\begin{tabular}{l c c c c c c c c c} 
\hline\hline 
~~ &$k=0$ & 1 & 2 & 4 & 8 & 16 & 32 & 64 & 128 \\ [0.5ex] 
\hline 
simul.& 1.33 & 1.53~ & 1.53~ & 1.56~ & 1.63~ & 1.69~ & 1.76~ & 1.79~ & 1.83~\\ 
analytic~~& 1.33 & 1.50~ & 1.53~ & 1.61~ & 1.69~ & 1.76~ & 1.82~ & 1.86~ & 1.90~\\ [1ex] 
\hline 
\end{tabular}
\label{table:exp} 
\end{table}

We introduced the attribute of frugality into a prototypical random-walk
foraging model.  Frugality means that a forager does not consume food until
it has reached a partially depleted state in which it can survive only $k$
additional time units without food before starving.  The interplay between
\emph{conservation} (not eating) and \emph{consumption} leads to a rich
dynamics in which the lifetime of the forager is maximized at an optimal
level of frugality.  While it naively seems that being frugal is inherently
risky, this strategy turns out to be superior to that of the normal forager,
which always eats when it encounters food.  We also extended our approach to
obtain the forager lifetime in any dimension and for general frugality
threshold $k\ll\mathcal{S}$ by exploiting the classic formalism for visits to
distinct sites of a random walk.

Finally, the super exponential decay of the survival probability in
Eq.~\eqref{lnSm} for $d\leq 2$ has important implications for the
self-avoiding flight problem.  For the classic self-avoiding walk (SAW) (see,
e.g.,~\cite{madras1996self}), the survival probability---the ratio of the
number of $n$-step SAWs to $n$-step random walks on the same lattice---decays
exponentially with $n$.  The non-exponential behavior in \eqref{general}
suggests that the self-avoiding flight could lie in a separate universality
class.

We acknowledge support from the European Research Council Starting Grant No.\
FPTOpt-277998 (OB), a University of California Merced postdoctoral fellowship
(UB), grant No.~DMR-1608211 from the National Science Foundation (UB and SR),
and the John Templeton Foundation (SR).

\bibliographystyle{apsrev4-1}

%

\end{document}